\documentclass[conference]{IEEEtran}
\IEEEoverridecommandlockouts
\usepackage{cite}
\usepackage{amsmath,amssymb,amsfonts}
\usepackage{graphicx}
\usepackage{textcomp}
\usepackage{xcolor}

\usepackage[utf8]{inputenc}
\usepackage{booktabs} % For formal tables
\usepackage{amsmath}
\usepackage{graphicx}
\usepackage{multirow}
\usepackage[noend]{algorithm2e}
\usepackage{tabularx}
\usepackage{tabulary}
\usepackage{cleveref}
\usepackage{float}
\usepackage{makecell}
\usepackage{breakcites}
\usepackage{optidef}
\usepackage{microtype}
\usepackage{etoolbox}
\usepackage{algpseudocode} 
\usepackage{lipsum}
\usepackage{array}
\usepackage{siunitx}
\usepackage{pgfplots}
\usepackage[caption=false,font=normalsize,labelfont=sf,textfont=sf]{subfig}
\pgfplotsset{width=8cm,compat=1.9}
\newcolumntype{P}[1]{>{\centering\arraybackslash}p{#1}}
\def\BibTeX{{\rm B\kern-.05em{\sc i\kern-.025em b}\kern-.08em
    T\kern-.1667em\lower.7ex\hbox{E}\kern-.125emX}}
\usepackage{makecell}

\makeatletter
\newcommand{\printfootsymbol}[1]{%
  \textsuperscript{\@fnsymbol{#1}}%
}
\makeatother

\makeatletter
\newcommand{\printfnsymbol}[1]{%
  \textsuperscript{#1}%
}
\makeatother

\makeatletter
\newcommand{\printsnsymbol}[1]{%
  \textsuperscript{2}%
}
\makeatother

\makeatletter
\newcommand{\printtnsymbol}[1]{%
  \textsuperscript{3}%
}
\makeatother

\begin{document}

\title{Simulation of Real-time Routing for UAS traffic Management with Communication and Airspace Safety Considerations}

\author{\IEEEauthorblockN{
Zhao Jin\printfnsymbol{1}\printfootsymbol{1}\thanks{\printfootsymbol{1}These authors contributed equally.}, 
Ziyi Zhao\printfnsymbol{1}\printfootsymbol{1}, 
Chen Luo\printfnsymbol{1}, \\
Franco Basti\printtnsymbol{3}, 
Adrian Solomon\printtnsymbol{3}, 
M. Cenk Gursoy\printfnsymbol{1}, 
Carlos Caicedo\printsnsymbol{2}, 
Qinru Qiu\printfnsymbol{1}}
\IEEEauthorblockA{
\textit{\printfnsymbol{1} Department of Electrical Engineering \& Computer Science}, \textit{Syracuse University}, Syracuse, NY 13244, USA \\
\textit{\printsnsymbol{2} School of Information Studies}, \textit{Syracuse University}, Syracuse, NY 13244, USA \\
\textit{\printtnsymbol{3} Thales Digital Aviation Customer Success and Innovation}, \textit{Thales USA}, Arlington, VA 22202, USA \\
\{zjin04, zzhao37, cluo05, mcgursoy, qiqiu\}@syr.edu, ccaicedo@syr.edu, \{franco.basti.e, Adrian.solomon.e\}@thalesdigital.io}
}

\maketitle

\begin{abstract}
Small Unmanned Aircraft Systems (sUAS) will be an important component of the smart city and intelligent transportation environments of the near future. The demand for sUAS related applications, such as commercial delivery and land surveying, is expected to grow rapidly in next few years. In general, sUAS traffic routing and management functions are needed to coordinate the launching of sUAS from different launch sites and determine their trajectories to avoid conflict while considering several other constraints such as expected arrival time, minimum flight energy, and availability of communication resources. However, as the airborne sUAS density grows in a certain area, it is difficult to foresee the potential airspace and communications resource conflicts and make immediate decisions to avoid them. To address this challenge, we present a temporal and spatial routing  algorithm and simulation platform for sUAS trajectory management in a high density urban area that plans sUAS movements in a spatial and temporal maze taking into account obstacles that are either static or dynamic in time. The routing allows the sUAS to avoid static no-fly areas (i.e. static obstacles) or other in-flight sUAS and areas that have congested communication resources (i.e. dynamic obstacles). The algorithm is evaluated using an agent-based simulation platform. The simulation results show that the proposed algorithm outperforms other route management algorithms in many areas, especially in processing speed and memory efficiency. Detailed comparisons are provided for the sUAS flight time, the overall throughput, conflict rate and communication resource utilization. The results demonstrate that our proposed algorithm can be used to address the airspace and communication resource utilization needs for a next generation smart city and smart transportation.
\end{abstract}


\section{Introduction}\label{sec:introduction} 

It is foreseeable that the emerging technology of Unmanned Aircraft Systems (UAS) will enable many new applications such as package delivery, rescue mission, senior assistance, etc. The traffic demand from new entrants in low-altitude airspace is forecasted to be orders of magnitude far greater than existing commercial aviation. Demands for controlling and monitoring this airspace will increase in particular in large, metropolitan areas. The safe and efficient operation of the UAS has two minimum requirements, a set of disjoint flight trajectories that ensures the minimum distance between the small UAS (sUAS) and a reliable wireless communication channel that guarantees the exchange of status report and command and control information between the sUAS and ground control stations (GCSs).

Cellular-connected sUAS communications, where the sUAS are supported by base stations as new aerial users, has recently attracted attention from the research community \cite{zhang2018cellular}, \cite{azari2019cellular}. Due to the availability of the existing infrastructure and radpid development of 5G technology, this becomes a promising solution for UAS communication. However, these studies generally consider small number of  sUAS (e.g., a single sUAS in \cite{zhang2018cellular}) and do not address air space management. We consider both air space and communication network capacity as resources. The demands and utilizations of these two resources in a sUAS system are highly correlated, and hence should be managed together.

In this paper, we present a temporal and spatial (T-S) routing algorithm for sUAS trajectory planning. The algorithm helps in the management of the air traffic of a metropolitan area that has high sUAS densities by assisting in the planning of each sUAS trajectory in advance. The centralized management ensures sUAS safety by proactively avoiding conflicts while ensuring the availability of communication resources. The goal is to find the flight trajectory of an sUAS that minimizes the flight distance while satisfies the air space and communication resource constraints. What determines the control thrust of the sUAS flight usually is not its flight distance but its accumulated velocity acceleration. We further improve the algorithm to minimize control thrust by searching for trajectories that has minimum number of turns. The routing algorithm will be applied before the launch of every sUAS. It returns an energy efficient trajectory from source to destination that is collision free and has guaranteed connectivity. If no trajectory that satisfies the constraint can be found, then the sUAS will not be launched.

Using a multi-agent air traffic resource usage simulator (MATRUS) \cite{zhao2019simulation}, we compare the performance of our proposed algorithm to other methods. Simulation results show that, compared to the scenario without traffic management, our proposed UAS routing algorithm completely eliminate the potential conflict while maintain a 100\% connectivity during the mission with 2~3.3\% reduction in throughput and less than 2.74\% increase in flight time(Table~\ref{tab1:algorithm_coparison}).

The sUAS  communicate with ground control stations (i.e. a command and the control center) through the cellular network, hence the availability of the communication resources will impact the decisions related to sUAS traffic management. The effectiveness of such management will eventually determine the scale of sUAS applications/services that can be supported by the given air space and existing communication/monitoring infrastructure, and also shape the planning of future infrastructure deployment. The UAS traffic management system must be fully autonomous, so that it can handle a large number of sUAS simultaneously and continuously.

The rest of the paper is arranged as follows: In Section~\ref{sec:related}, we review related work in UAS route management and conflict avoidance. This is followed in Section~\ref{sec:algorithm} by details about the proposed algorithm. Section~\ref{sec:experiment} describes our experiments and evaluation of the results. Finally, Section~\ref{sec:conclusion} summarizes this work.

\section{Related Work}\label{sec:related} 

Over the past 5 years, sUAS have played an increasingly critical role in many fields \cite{puri2005survey}. With the rise in popularity of sUAS, many important and notable issues regarding sUAS traffic management have been discovered. Therefore, numerous methods and paradigms have been proposed to solve these issues. These proposed methods can be divided into two main categories. One category focuses on the centralized scheduling and management of multiple sUAS, via unmanned air system traffic management systems \cite{kopardekar2016unmanned}\cite{sastry1995hybrid}. We refer to these approaches as centralized control. The other category focuses on the actions of a single sUAS, such as obstacle and collision avoidance \cite{albaker2009survey}, and is referred to as distributed control.

Much of the existing UAS trajectory generation research focuses on the generation of trajectories for a single sUAS that are energy efficient and stable. Constraints such as obstacle avoidance and rigid body dynamics are considered. Some of the classical approaches apply rapidly-exploring random trees \cite{lavalle1998rapidly} and Voronoi graphs \cite{bortoff2000path} \cite{tisdale2009autonomous}. In \cite{odelga2016obstacle}, the author presented an indoor algorithm to navigate sUAS to avoid collisions. By importing geometrical constraints, \cite{thanh2018completion} proposed a solution to avoid collisions in a static environment. In \cite{wang2015obstacle}, to improve obstacle avoidance for sUAS, a method based on optical flow is presented. Others use machine learning approaches. For example, \cite{li2019autonomous}\cite{eslamiatautonomous} developed a deep reinforcement learning framework that learns how to perform energy efficient waypoint planning. However, those works assume static obstacles and a single sUAS.

In addition, multiple sUAS trajectory planning has been studied as a multi-agent cooperative system and solved in a rolling horizon approach using dynamic programming \cite{beard2003multiple} or mixed integer linear programming \cite{song2016rolling}. Another strategy involves setting an artificial reactive field around each UAS \cite{rimon1992exact}. However, these approaches do not consider any additional resources other than the airspace. The availability of the communication resource has not been integrated as a constraint into these frameworks.

Most of the existing trajectory planning algorithms consider continuous Euclidean space and the sUAS can have an arbitrary trajectory as long as certain constraints are satisfied. Hence a closed-form representation of the trajectory can be obtained. Although this may allow us to find simple analytical optimal solutions, it leads to unstructured trajectories. When the number of sUAS increases, such irregularity leads to a traffic pattern that is unpredictable. Furthermore, with a large number of sUAS, to describe all constraints (i.e. collision avoidance) in closed-form and solve the optimization problem analytically is almost impossible. Recently, a very strict and rigid airspace structure to handle dense operation in the urban low altitude environment was suggested by work on Unmanned Aircraft System (UAS) Traffic Management (UTM) at NASA in \cite{jang2017concepts}. The author explored UAS operations in non-segregated airspace and managed the risk of mid-air collision to a level deemed acceptable to regulators. In the paper, the airspace is divided into multiple layers, and the layers are further divided into orthogonal sky lanes. However it focused only on defining the regulations of the UTM system instead of solving any optimization problems. 

In this work, we focused on optimizing the computed trajectory for sUAS in order to prevent collisions, achieve shortest distance, and at the same time meet other realistic environmental constraints \cite{ferguson2005guide}. These constraints include avoiding no-fly zones or restricted areas, avoiding areas that do not have cellular signal coverage or that are temporarily experiencing congestion in the cellular network, etc. These constraints will be modeled as static and/or dynamic obstacles in the airspace, and can be considered simultaneously as part of the airspace management environment.

\section{Airspace and Communication Aware Routing}\label{sec:algorithm} 

\subsection{Environment Assumptions}

One of the main objectives of sUAS traffic management is to maximize the throughput while avoiding any potential conflicts. The conflict is defined as the situation in which the distance between two sUASs is smaller than the given threshold. We divide deconfliction techniques into two categories, reactive and proactive. A sUAS with reactive deconfliction capabilities perceives an imminent conflict and adjusts its trajectory locally to avoid it. The conflict could be detected via on-board sensors, or through communication with nearby sUAS and the control center. In either case, to ensure safety operation, the sUAS needs to have a high amount of computing power to respond in a short time and avoid the conflict. Furthermore, reactive deconfliction leads to unpredictable traffic patterns. During trajectory adjustment, a sUAS not only needs to consider the upcoming conflict, but also any potential new conflict that may be caused by the changing of its current trajectory. In a high density area, this problem will soon become intractable. This has been confirmed by the works in \cite{kopardekar2014unmanned}\cite{clothier2015structuring}, which stated the importance of architecting a UTM solution capable of handling high UAS traffic demand and that in some situations free flight operations with fully decentralized trajectory planning are not feasible or will result in very inefficient airspace operations.

The proactive deconfliction technique plans a conflict free trajectory for each sUAS at launch time or at the time when it enters controlled airspace. Because the control center has the trajectory information of all sUAS in a designated airspace, it can easily find a conflict free path for the incoming sUAS if such path exists. If such path cannot be found, the launch of the sUAS will be delayed or the sUAS cannot enter the airspace until a path is available. Although the routing procedure may have high complexity, it is done in the control center, hence energy or computing resources will not be a limiting factor. Other air traffic constraints, such as no-fly zones can easily be integrated into the routing procedure.

With a large number of sUAS in a given airspace area, if the sUAS trajectory has the freedom of taking any angle at any speed, the routing will be extremely difficult as the search space is infinite. Some constraints on the sUAS trajectory must be imposed to reduce the route planning search space. In this work, we adopt the “sky lane” concept proposed by the NASA UTM group \cite{jang2017concepts} and limit the trajectory to be a Manhattan style trajectory, i.e. the sUAS can only make 90 degree turns, and they fly at constant speed. To improve the predictability and to ease the de-conflict cost. Similar to \cite{jang2017concepts}, we divide the airspace using a grid pattern. The size of the grid cell is defined by the minimum separation distance between UASs for safe operations. It is guaranteed that UASs will be conflict-free if they travel along the center of each grid cell.

\begin{figure}[htbp]
\centering
\captionsetup{justification=centering,margin=1.5cm}
\includegraphics[width=2.5in]{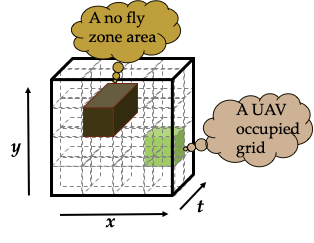}
\caption{3 Dimensional Spatial-Temporal Environment}
\label{fig:3d_environment}
\vspace{-1em}
\end{figure}

To simplify discussion and illustration, we assume that all sUASs fly at the same height and our search space consists of only 3 dimensions: $x$, $y$ and $t$ (i.e. two dimensions for space and one for time). The entire 3 dimensional space is divided into equal sized grids as shown in Fig.~\ref{fig:3d_environment}. The size of each grid cell is ($W$, $W$, $\delta$), where $W$ is the minimum distance between sUASs which ensures that they are not in conflict. The amount of time that a UAS needs to travel a distance $W$ is denoted as $\delta$. A function $M(x, y, t)\rightarrow\{-1, 0\}$ maps each grid to a label, where "-1" represents an occupied grid cell and “0” represents unoccupied grid cell. As we can see, if location $(x, y)$ belongs to a no-fly zone, then $M(x, y, t) = -1$, 0$\leq$ $t$ $\leq$ $\infty$. If a UAS flies through a location $(x, y)$ at time $t_1$, then $M(x, y, t_1)=-1$. A UAS trajectory starts from a source grid cell $(x_s, y_s, t_s)$, and end at any one of the destination grid cells, $(x_d, y_d, t_d)$, where $(x_s, y_s)$ and $(x_d, y_d)$ are the coordinates of the flight source and destination, respectively, and $t_s,t_d$ are the flight start time and arrival deadline. We refer to this grid system as the temporal-spatial (T-S) maze.

Traditional maze routing has been widely used in electronic design automation to route the on-board or on-chip interconnects. Two interconnects cannot occupy the same grid cell, otherwise it will cause a short circuit. However, by using the $t$ axis, two sUASs can occupy the same space as long as they are there at different times. Hence, the trajectory search space should consist of three dimensions, $x$, $y$ and $t$. Two sUASs are conflict free, if their trajectory has no intersection in the multi-dimensional spatial and temporal space. In a T-S maze, obviously, any route must move towards the direction where $t$ increases.

\subsection{Baseline T-S Routing Algorithm}

In this paper, the baseline method we selected is a Breadth First Search (BFS) based maze routing algorithm. It consists of two stages, flooding and traceback.
% , as shown in Fig.~\ref{fig:bfs}. 
The flooding stage essentially performs a breadth first search. Starting from the source grid, every non-occupied grid is labeled by its Manhattan distance from the source grid cell using BFS until one of the destination grid cells is reached. Then following the descending order of the labels, a path is traced back from the destination to the source. In order to reduce the control thrust during the flight, the grid cell that is in the same direction (towards the source) as the previous move will be picked with higher priority during the traceback to reduce the number of turns in the trajectory. The complete algorithm is given in Algorithm~\ref{alg:the_BFS}.

%BFS Scheduling Pseudo Code
\begin{algorithm}
    \SetKwInOut{Input}{Input}
    \SetKwInOut{Output}{Output}
    \underline{T-S Routing} $(E(x',y',t'), s_x, s_y, s_t, d_x, d_y)$\;
    \Input{Environment Matrix, Start Coordinates($s_x$, $s_y$, $s_t$), Destination Coordinates($d_x$, $d_y$)}
    \Output{The optimal routing trajectory of one UAS}
    \texttt{\\}
    Queue $Q$\;
    $Q$.enqueue$(start\_position)$\;
    $mark$ $start\_position$ $as$ $VISITED$\;
    \While{$Q \neq \emptyset$}
    {
        $node$ =  $Q$.poll()\;
        \If{$node == destination$}
        {
            $set$ $DestinationTimeStep$\;
            break\;
        }
        \ForEach{$neighbor \in Neighbors(node)$}
        {
            \If{$neighbor$ $IS\_VALID$}
            {   
                $mark$ $neighbor$ $as$ $VISITED$\;
                $Q$.enqueue$(neighbor)$\;
            }
        }
    }
    \If{$DestinationTimeStep$ $IS\_EXIST$}
    {
        $get$ $trajectory$ $by$ $TRACEBACK$\;
    }
    \ForEach{$position \in trajectory$}
    {
        $mark$ $E[position.x][position.y][position.t]$ $as$ $OBSTACLE$\;
    }
    \Return $trajectory$\;
    
    \texttt{\\}
    \caption{T-S Routing Algorithm}
    \label{alg:the_BFS}
\end{algorithm}

\subsection{Sparse Represented Temporal-Spatial (SRTS)  Routing}

In the worst case, the BFS routing needs to label all grids in the 3D T-S maze to reach the destination. A naive implementation has the space complexity  $O(X\times Y \times T)$ , where X, Y and T are the maximum dimensions of the airspace and the maximum time that the air traffic will last. Its memory and computation complexity is prohibitively high. To improve the routing computation speed for real-time applications, we propose a Sparse Represented TS routing (SRTS) procedure based on the A$\ast$ routing algorithm \cite{seet2004star}\cite{nilsson2009quest}. 

The UAS route planning problem has to consider two types of obstacles, static and dynamic. The static obstacles refer to the invariant geographical constraints in the route planning area. The dynamic obstacles represent the time variant constraints.

Unlike original T-S routing,  our routing method uses a 2D map with dimensions $X\times Y$. (This can be extended to a 3D map if the UASs fly in different altitudes.) The obstacles are divided into 2 categories. The static obstacles are projected onto a 2D map, each of the obstacles occupies a specific location. Since the static obstacles are time invariant, the information along the $t$ axis is redundant and hence it can be eliminated. For the dynamic obstacles, we exploit their spatial sparsity and store them using hash tables along the $t$ axis. Each location on the $t$ axis is associated with a hash table, which stores the $(x, y)$ coordinates of dynamic obstacles at the corresponding time.  Using an instant refreshing mechanism, the SRTS routing algorithm only stores dynamic obstacles at or beyond the current time step. All the obstacles in the prior time will not affect the trajectory planning of the current sUAS, hence will be removed automatically. In this way, the 3D environment considered in the original T-S routing is compressed into a set of sparsely represented 2D points corresponding to dynamic obstacles sampled at each time step from the present up to a future time $T’$. The value of $T’$ is determined by the longest flight time for UASs currently in the air or about to launch. The experimental results show that the instant refreshing mechanism can significantly reduce the demand of the memory resources. 

Using the new environment representation, the routing algorithm needs to check both 2D static obstacles and 3D dynamic obstacles to acquire next moveable neighbor cell. Only the neighbor cell which has no conflict with either type of obstacle will be selected for the next potential movements, as shown in Algorithm~\ref{alg:the_candidate_selection}. 

%Candidate Selection Pseudo Code
\begin{algorithm}
    \SetKwInOut{Input}{Input}
    \SetKwInOut{Output}{Output}
    \underline{Candidate Selection} $(currNode,  ClosedList)$\;
    \Input{Current Position(currNode), Past Selected Positions(ClosedList)}
    \Output{The candidate neighbors of current position}
    \texttt{\\}
    $neighbors$= $\emptyset$\;
    \ForEach {$position$ in $Directions$}
    {
        $check$ $position$ $in$ $2D$ $static$ $projection$\;
        $check$ $position$ $in$ $3D$ $dynamic$ $projection$\;
        $CheckSignalStrength$\;
        \If{$position$ $IS\_VALID$ \& \\ 
        $no$ $obstacle$ $in$ $2D/3D$ $projections$ \& \\
        $have$ $enough$ $signal$ $support$ \& \\
        $position \notin ClosedList$ }
        {
            $position$ $initialization$\;
            $neighbors$.add$(position)$\;
        }
    }
    \Return $neighbors$\;
    \texttt{\\}
    \caption{Candidate Selection Algorithm}
    \label{alg:the_candidate_selection}
\end{algorithm}

\subsection{Routing for Connectivity}\label{connectivity}

After checking the airspace resources, the algorithm also needs to check the availability of communication resources in order to decide the next move in the trajectory. In this work, a simplified communication model is used, where each base station has $N$ orthogonal communication channels and each channel can serve only one sUAS. At anytime, a sUAS will connect to one base station through one of the channels that are available at that base station. We adopt the following simple log-distance path loss model:
\begin{equation}
    PL(d) = PL(d_{0}) + 10n\log_{10}(\frac{d}{d_{0}}) + x, d_{f}\leq d_{0}\leq d
    \label{eqn:signal}
\end{equation}

where $PL(d)$ is the path loss in dB at distance $d$ and $PL(d_0)$ is the path loss in dB at a reference distance $d_0$, $x$ is a Gaussian distributed random variable, however for simplicity, it is set to 0 in this work. The parameter $n$ is the path loss exponent, which is set to 3 for line-of-sight links and 3.5 for non-line-of-sight. Since the sUAS connect to base stations from a high altitude, the connection has a higher chance to be line-of-sight. In this work we set the line-of-sight probability to be 0.9 and the non-line-of-sight probability to be 0.1. 

We assume that a link can be established between a sUAS and a base station if the signal loss of the path between them is less than a given threshold. Otherwise, the communication link cannot be established. At any given time, the sUAS only connects to the base station that has the highest signal strength through an available channel. We assume that the connection between sUAS and base station is dynamic and we ignore the time and cost of the handover process.

Algorithm~\ref{alg:the_check_signal_strength} shows the procedure for communication resource availability checking. First, the communication  resources (i.e. channels) will be allocated to the sUASs that are already in the air to give them higher priority than the sUAS that are about to launch. After updating the base station’s list of available resources, the distance between the potential new location of the UAS after a movement step and each base station will  be calculated and the potential  signal loss will be estimated. At the end, the movement step will be considered as a valid step if  a base station with  an available communication channel and acceptable link quality can be found. Otherwise, the movement step will not be considered for the current position of the UAS. 

During the route planning, when checking for communication resource availability, we assume the connections between sUAS and the base stations are line-of-sight. This may not always be the case if there are obstacles present in the environment. Therefore, the estimated signal strength during the routing stage may not be the exact signal strength during the real flight. We refer to the former as the $inner$ $belief$ of the resource utilization and the latter as the ground truth. Our simulation results show that, under the simple channel model, the $inner$ $belief$ will be close to the ground truth.

%Candidate Signal Strength Pseudo Code
\begin{algorithm}
    \SetKwInOut{Input}{Input}
    \SetKwInOut{Output}{Output}
    \underline{Signal Strength Check} $(x, y, t)$\;
    \Input{Candidate Position(x, y, t)}
    \Output{Whether a position can build the communication link with a base station}
    \texttt{\\}
    $allocate$ $base$ $station$ $resources$ $for$ $in$-$flight$ $UASs_{(t)}$\;
    $sort$ $BaseStationCandidates$ $by$ $distance$ $in$ $ascending$ $order$\;
    \ForEach {$basestation$ in $BaseStationCandidates$}
    {
        
        $get$ $distance$ $between$ $UAS$ $and$ $basestation$\;
        $calculate$ $SignalLoss$\;
        \If{$basestation$ $has$ $available$ $channel$ \& \\ 
        $SignalLoss \leqslant LOSS\_THRESHOLD$}
        {
            $reset$ $base$ $station$ $resources$\; 
            \Return $TRUE$\;
        }
    }
    $reset$ $base$ $station$ $resources$\;
    \Return $FALSE$\;
    \texttt{\\}
    \caption{Signal Strength Check Algorithm}
    \label{alg:the_check_signal_strength}
\end{algorithm}

\subsection{Energy-aware Routing Constraint}
The sUAS onboard battery imposes a physical constraint that limits the flight time and the maximum range that a UAS can reach. Energy efficiency is an important factor to consider during sUAS trajectory management, because it allows the sUAS to serve more missions before battery recharge, and reduces the possibility of mission failure due to battery depletion.

Both the traditional T-S routing and the proposed SRTS routing aims at searching for the trajectories that have the minimum distance. For Manhattan style routing,  there will be multiple trajectories with equal distance that are minimum among all other trajectories, and they will be selected without any preference. However, minimum distance does not necessarily mean minimum control thrust or minimum energy dissipation. Extra control thrust is needed when a sUAS accelerates or de-accelerates. While the sUAS is assumed to fly at a constant speed,during most of the time, the change in velocity happens when the flight direction changes. In other words, each time a sUAS turns (in 90 degree), it will dissipate extra energy.

Consequently, in our SRTS algorithm, a new penalty is introduced and considered each time the trajectory changes its direction. The turning penalty will accumulate along current planning path. The routing algorithm is inclined to explore the path which has less turn to mitigate the penalty.  Meanwhile, the shortest path always has the highest priority among all trajectories with the same turning penalty.

\subsection{Overall SRTS Routing Algorithm}

The SRTS routing is performed on the 2D surface with the static obstacle information. The algorithm calculates two costs for each potential neighbor cell. The first cost is called the movement cost, which describes the expense of moving from the current position to the potential neighbor cell. For each position, the movement cost is accumulated, so it can relieve the oscillation between two adjacent positions. The other cost is called the destination cost, which stands for the expense of moving from the potential neighbor cell to the destination. The last cost is defined as the turning cost, which penalize the turns in the UAS planning path.The definition of movement and destination costs are adopted from the original A$\ast$ routing algorithm. We choose to calculate the costs based only on the $X$, $Y$ distances, because in general UAS flight energy increases as the travel distance increases. However, as a T-S routing algorithm, it is possible that the GCS will ask an UAS to stay at a specific location to wait to resolve the potential conflict. It will be part of our future work to incorporate flight time into the cost function. The pseudo code of the SRTS algorithm is given in the Algorithm~\ref{alg:the_ASTAR}.

%Static-Dynamic Scheduling Pseudo Code
\begin{algorithm}
    \SetKwInOut{Input}{Input}
    \SetKwInOut{Output}{Output}
    \underline{Sparse Represented TS Routing} $(s_x, s_y, s_t, d_x, d_y)$\;
    \Input{Start Coordinates($s_x$, $s_y$, $s_t$), Destination Coordinates($d_x$, $d_y$)}
    \Output{The optimal routing trajectory of one UAS}
    \texttt{\\}
    $OpenList$ = $\emptyset$\;
    $ClosedList$ = $\emptyset$\;
    $OpenList$.add($start\_position$)\;
    \While{$OpenList \neq \emptyset$}
    {
        $node$ = $OpenList$.poll()\;
        $\textbf{Instant}$ $\textbf{Refreshing}$ $\textbf{Mechanism}$
    
        \If{$node$ == $destination$}
        {
            $trajectory$ = retriveTrajectory$(node)$ \;
            $break$;
        }
        $ClosedList$.add$(node)$\;
        \ForEach{$neighbor \in Candidate Selection$}
        {
            \If{$neighbor$ $\in$ $OpenList$}
            {
                $neighbor$ = $OpenList$.get$(neighbor)$ % not accurate, need edit
            }
            $calculate$ $neighbor's$ $NewMovementCost$\;
            \If{$neighbor$ turns} 
            {
                $update$ $neighbor's$ $TurnPointCost$
            }
            \If{$NewMovementCost$ $<$ $OldMovementCost$ $||$ $neighbor \notin OpenList$ }
            {   
                $update$ $neighbor's$ $MovementCost$\;
                $update$ $neighbor's$ $DestinationCost$\;
                $OverallCost$ = $MovementCost$ + $DestinationCost$ + $TurnPointCost$\;
            } 
            \If{$neighbor \notin OpenList$}
            {
                $OpenList$.add($neighbor$)\;
            }
        }
    }
    \ForEach{$position \in trajectory$}
    {
        $mark$ $OBSTACLE$ $in$ $3D$ $dynamic$ $projection$\;
    }
    \Return $trajectory$\;
    \texttt{\\}
    \caption{Sparse Represented TS Routing Algorithm}
    \label{alg:the_ASTAR}

\end{algorithm}

\section{Experimental Results \& Analysis}\label{sec:experiment} 

We demonstrate the performance of the SRTS algorithm in UAS trajectory planning on the MATRUS simulation framework \cite{zhao2019simulation}. The environment settings in MATRUS contain parameters that specify the base station configurations, individual UAS action configurations and air space configurations.

In this paper, the number of base stations and the channels available for UAS communications at each base station are fixed to be 10 and 8, respectively. For each UAS, the trajectory mode is set to be Manhattan style. The UAS mission generation interval varies  from 10, 20, and 30 seconds. They will be referred to as the high, medium and low traffic configurations, respectively, in the rest of this section. And, the number of no-fly zones is either none or 2. Each combination of parameters defines one specific scenario. The reported results is the average of 10 runs for each specific scenario. For each run of one scenario, the simulation time is 20,000 time steps which corresponds to 20,000 seconds. Based on the observations from the experiments, the UAS simulation behavior becomes stable after 300 time steps. Therefore, the simulation length is sufficient for us to analyze different scenarios.

Four sUAS launching areas and four landing areas are distributed in the map, which is 90 square miles in size. Their locations are selected based on the distribution of business and residential areas in Upstate, New York. Each launching area has a different launching probability. In a given interval, each launching area will request to launch a sUAS with a given probability. For each launch request, the landing area is randomly selected from the four candidates. Given the launch and landing areas, the exact launch and landing spots are randomly picked within the areas. 

\begin{figure}[htbp]
\centering
\captionsetup{justification=centering,margin=1.5cm}
\includegraphics[width=3.0in]{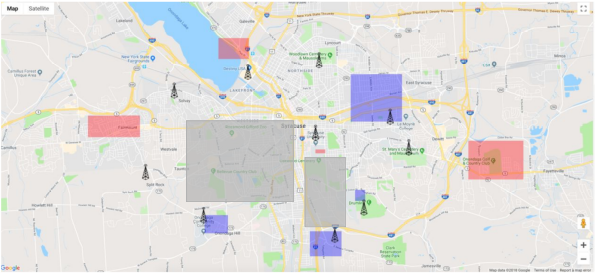}
\caption{Traffic Scenario Environment}
\label{fig:environment}
\vspace{-1em}
\end{figure}

One of the traffic scenario environments is illustrated in Fig.~\ref{fig:environment}. In the figure, the red rectangles represent the sUAS launching areas, the blue rectangles stand for the sUAS landing areas and the grey rectangles indicate blocked areas (i.e. the no-fly zones). The location of 10 base stations are also marked on the map. The coordinates of those base stations are set based on the actual base station facilities registered with the FCC.

\subsection{Evaluation Metrics}

Three metrics have been introduced to evaluate the performance of the routing algorithm: Average Throughput, Average Flight Time and Average Conflict Ratio.

To ensure flight safety, two flying sUAS must be separated by a sufficient distance. Because the current sUAS cannot make sharp turns or slow down immediately to a stop, leaving enough space for each sUAS is necessary for safety considerations. Therefore, centered at every location $(x, y)$ of the map, a square is drawn whose dimension is equal to the minimum separation distance. If a square is occupied by more than one sUAS at the same time, then location $(x, y)$ has a conflict at this particular time. By default, we use 18 meters as the minimum separation distance in the experiment. 

In our evaluation, the conflict ratio is used to analyze the safety metric. It is defined as the number of missions that have encountered at least one conflict during the flight divided by the total number of launched sUAS missions.

The sUAS throughput indicates the capacity of the simulated air space. It is measured by the number of launched sUAS during a fixed time. There is a fundamental trade-off between safety and throughput. The trajectory planning algorithm can significantly reduce the conflict ratio, however it will also affect the throughput. The goal of the routing algorithm is to achieve maximum throughput while avoiding any potential conflicts.

Besides the average throughput of the entire simulated area, the performance of every single sUAS is also crucial. In this paper. The average flight time of individual sUAS has been considered as the last metric to evaluate the performance of the routing algorithm. In general, a longer average flight time indicates more detours during the flight and higher energy consumption. Hence, a viable routing algorithm should not lead to a large increase in the sUAS flight time.

\subsection{Conflict Elimination}

\begin{figure}[htbp]
\centering
\captionsetup{justification=centering,margin=1.5cm}
\includegraphics[width=3.4in]{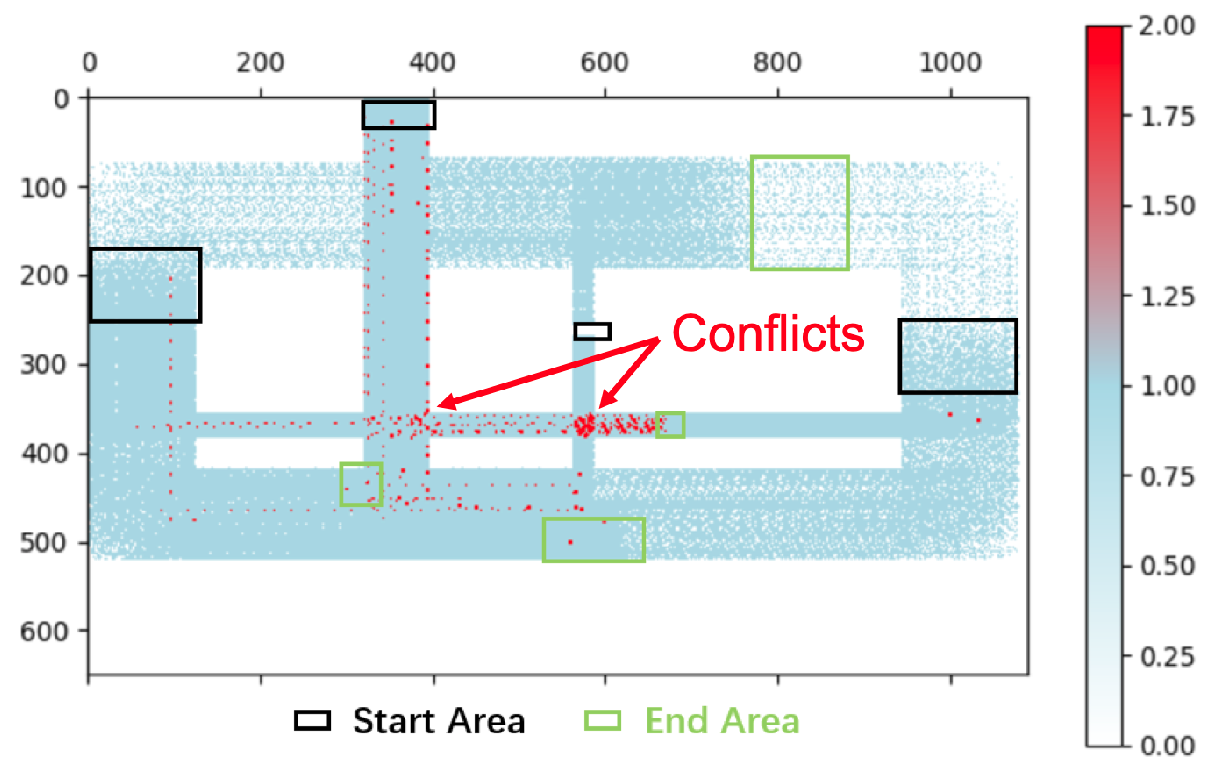}
\caption{sUAS Trajectory Density without Traffic Management}
\label{fig:no_routing}
\vspace{-1em}
\end{figure}

\begin{figure}[htbp]
\centering
\captionsetup{justification=centering,margin=1.5cm}
\includegraphics[width=3.4in]{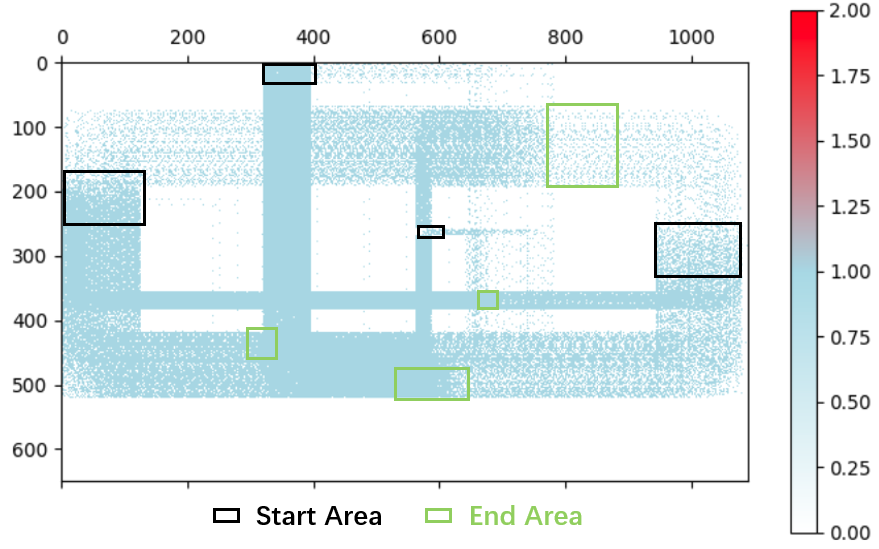}
\caption{sUAS Trajectory Density with Traffic Management (T-S Routing)}
\label{fig:bfs_routing}
\vspace{-1em}
\end{figure}

\begin{figure}[htbp]
\centering
\captionsetup{justification=centering,margin=1.5cm}
\includegraphics[width=3.4in]{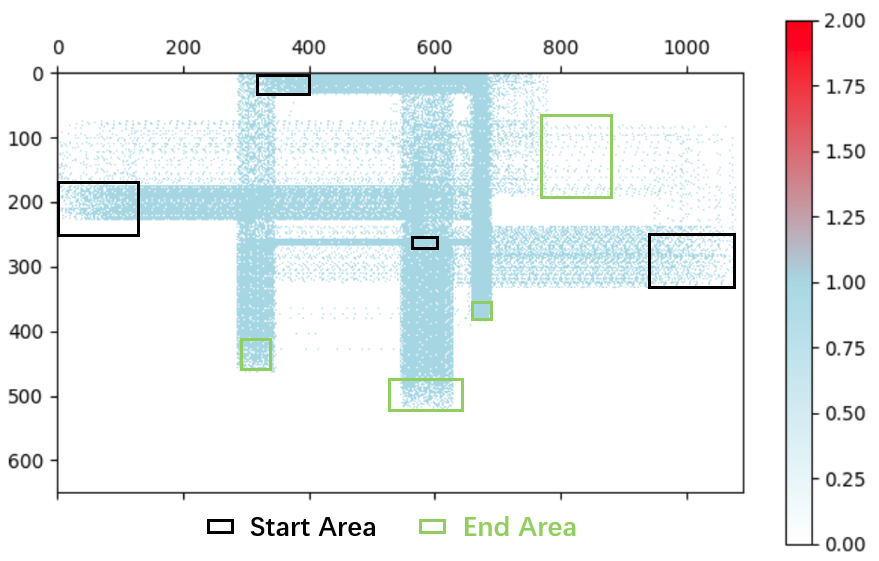}
\caption{sUAS Trajectory Density with Traffic Management (SRTS Routing)}
\label{fig:astar_routing}
\vspace{-1em}
\end{figure}

In the first experiment, we compare the maximum sUAS density in different air traffic scenarios, and demonstrate the effectiveness of our proposed routing algorithm. We visualize the distribution of maximum density of sUASs per grid location for the simulation scenario of high traffic without no-fly zones in Figs.~\ref{fig:no_routing},~\ref{fig:bfs_routing} and ~\ref{fig:astar_routing}. The black boxes in the density maps represent the sUAS launching areas and the green boxes stand for the sUAS landing areas. In the density map, the light blue spots indicate normal traffic density, i.e. the maximum density of sUAS in that area is 1 sUAS per grid cell. In contrast, the bright red spot indicates conflict, i.e. the maximum sUAS density is equal to or greater than 2 in the specific location. The white  areas are those where no sUAS has ever visited. Hence, the maximum density also represents the distribution of the sUAS trajectory. If a sUAS passes through the blocked area, it will be considered as a conflict. Table~\ref{tab1:algorithm_coparison} compares throughput, flight time and conflict rate for traffic scenarios without routing, with T-S routing, and with SRTS routing.

\begin{table*}[htbp]
\caption{Routing Algorithm Comparison}
\label{tab1:algorithm_coparison}
\centering
\begin{tabular}{|c|*{7}{c|}}
\hline
\multirow{2}{*}{\bf Traffic Type }  &
\multicolumn{2}{c|}{\bf Avg. Throughput} &
\multicolumn{2}{c|}{\bf Avg. Flight Time} &
\multicolumn{2}{c|}{\bf Avg. Conflict Ratio} \\
 & 0 No-Fly Zone & 2 No-Fly Zones & 0 No-Fly Zone & 2 No-Fly Zones & 0 No-Fly Zone & 2 No-Fly Zones
\\ \hline 
\multicolumn{7}{|c|}{\bf Heavy traffic (generation/10s) }\\
\hline
No Routing & 4006 & 4006 & 491.65s & 491.65s & 21.73\% & 46.10\%\\
\hline 
Baseline T-S Routing & 3897 (-2.72\%) & 3874 (-3.30\%) & 495.23s (+0.73\%) & 505.10s (+2.74\%) & 0.0\% & 0.0\%\\
\hline 
SRTS Routing & 3901 (-2.62\%) & 3880 (-3.15\%) & 494.79s (+0.64\%) & 504.67s (+2.65\%) & 0.0\% & 0.0\%\\
\hline
\multicolumn{7}{|c|}{\bf Medium Traffic (generation/20s)  }\\
\hline
No Routing & 1985 & 1985 & 491.73s & 491.73s & 11.53\% & 34.92\% \\
\hline 
Baseline T-S Routing & 1937 (-2.42\%) & 1930 (-2.77\%) & 494.86s (+0.64\%) & 503.52s (+2.40\%) & 0.0\% & 0.0\%\\
\hline 
SRTS Routing & 1940 (-2.27\%) & 1934 (-2.57\%) & 494.74s (+0.61\%) & 503.76s (+2.45\%) & 0.0\% & 0.0\%\\
\hline 
\multicolumn{7}{|c|}{\bf Light Traffic (generation/30s)  }\\
\hline
No Routing & 1315 & 1315 & 490.82s & 490.82s & 7.96\% & 31.8\%\\
\hline 
Baseline T-S Routing & 1288 (-2.05\%) & 1286 (-2.21\%) & 493.72s (+0.59\%) & 502.45s (+2.37\%) & 0.0\% & 0.0\%\\
\hline 
SRTS Routing & 1289 (-2.00\%) & 1288 (-2.05\%) & 493.85s (+0.62\%) & 502.32s (+2.34\%) & 0.0\% & 0.0\%\\
\hline 
\end{tabular}
\end{table*}

The first thing we can observe from Table~\ref{tab1:algorithm_coparison} is that, in the heavy traffic scenario, without trajectory management, 21.73\% missions will have conflicts. The conflict ratio has further increased to 46.10\% if no-fly zone conflicts are also considered. By applying the traffic management, both the original T-S routing algorithm and the SRTS routing algorithm can eliminate all the conflicts. The cost is 2.2$\sim$3.3\% reduction in throughput and less than 2.74\% increase in the flight time. The reason of the throughput reduction is because, with trajectory management, the sUAS that cannott find a conflict free path will not be launched. Therefore, the more restrictive the constraints are, the fewer sUAS that will be launched.

From Fig.~\ref{fig:no_routing} we can see that, without routing, some bright red points (i.e. conflicts) exist around  the center of map. From Fig.~\ref{fig:bfs_routing} and~\ref{fig:astar_routing} we can see that applying the TS routing and SRTS routing algorithm fully eliminate the conflicts. The sUAS trajectory concentrates in the upper part of the map in Fig.~\ref{fig:astar_routing}, because the candidate next move selection in SRTS follows a fixed priority, where going west or east always has higher priority than going north or south if all other conditions are the same.

\subsection{Communication Connectivity Improvement} 

\begin{figure}[htbp]
\centering
\captionsetup{justification=centering,margin=0.4cm}
\includegraphics[width=3.4in]{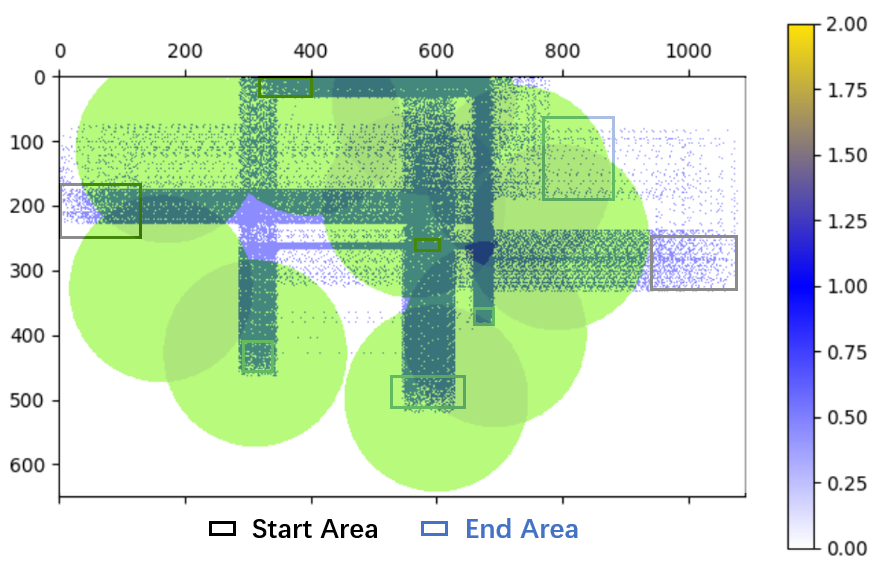}
\caption{Planned Trajectory without Connectivity Check Algorithm}
\label{fig:without_connectivity_check}
\vspace{-1em}
\end{figure}

\begin{figure}[htbp]
\centering
\captionsetup{justification=centering,margin=0.4cm}
\includegraphics[width=3.4in]{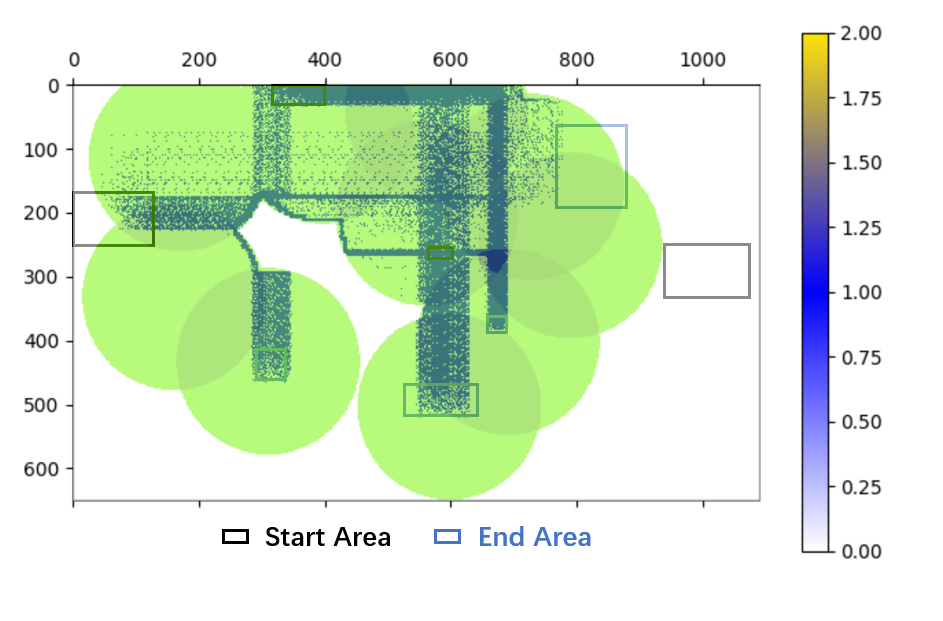}
\caption{Planned Trajectory with Connectivity Check Algorithm}
\label{fig:with_connectivity_check}
\vspace{-1em}
\end{figure}

In the second experiment, we demonstrate how SRTS routing can improve the connectivity of the sUAS with the cellular network. Using the log-distance path loss model given in Section~\ref{connectivity}, the UAS will establish a communication link with a base station if the path loss is less than 140dB. Otherwise, the communication link cannot be established. Table~\ref{tab1:connectivity} compares the routing results of the SRTS algorithm without consideration of connectivity (row 1) and with the consideration of connectivity (row 2). Fig.~\ref{fig:with_connectivity_check} and Fig.~\ref{fig:without_connectivity_check} show the sUAS trajectories with and without the connectivity check. The circles in the figure indicate the areas that are covered by a base station.

\begin{table}[htbp]
\captionsetup{justification=centering,margin=1cm}
\caption{The Comparison with Connectivity Check Algorithm}
\begin{center} 
\begin{tabular}{|P{2.2cm}|P{1.6cm}|P{1.5cm}|P{1.2cm}|}
\hline
\textbf{Traffic Type} & \textbf{Avg. In-flight UASs} & \textbf{Avg. Flight Time} & \textbf{No Link Rate} \\
\hline
\multicolumn{4}{|c|}{\bf Heavy Traffic (generation/10s)  }\\
\hline
SRTS Routing w/o. Conn. Check & \multirow{2}*{95.95} & \multirow{2}*{491.46s} & \multirow{2}*{85.15\%} \\
\cline{1-4}
SRTS Routing w. Conn. Check & \multirow{2}*{70.03} & \multirow{2}*{517.82s} & \multirow{2}*{0.00\%}  \\
\cline{1-4}
\multicolumn{4}{|c|}{\bf Medium Traffic (generation/20s)  }\\
\hline
SRTS Routing w/o. Conn. Check & \multirow{2}*{49.28} & \multirow{2}*{491.43s} & \multirow{2}*{43.48\%}  \\
\cline{1-4}
SRTS Routing w. Conn. Check & \multirow{2}*{41.44} & \multirow{2}*{513.12s} & \multirow{2}*{0.00\%}  \\
\cline{1-4}
\multicolumn{4}{|c|}{\bf Light Traffic (generation/30s)  }\\
\hline
SRTS Routing w/o. Conn. Check & \multirow{2}*{32.44} & \multirow{2}*{491.44s} & \multirow{2}*{26.25\%}  \\
\cline{1-4}
SRTS Routing w. Conn. Check & \multirow{2}*{28.66} & \multirow{2}*{512.82s} & \multirow{2}*{0.00\%}  \\
\cline{1-4}
\end{tabular}
\label{tab1:connectivity}
\end{center}
\end{table}

Our simulation results show that without checking the connectivity, in heavy traffic situation, the no link rate is 85.15\%. This means 85.15\% of sUAS will experience a certain period of time in its mission in which no cellular link can be established to communicate with the GCS. The no link rate reach 43.48\% and 26.25\% in medium and light traffic. Although the routing algorithm can plan a conflict free trajectory for those sUAS, some locations along the trajectory either do not have coverage from the cellular network (as shown in Fig.~\ref{fig:without_connectivity_check}) or the available channels have been depleted due to congestions. By applying the connectivity check, the no link rate is reduced to 0\%. From Fig.~\ref{fig:with_connectivity_check} we can see that the sUAS only fly in the areas which are covered by the base stations. We can also observe that with the connectivity check, the average number of  sUASs in the air is decreased by 27\%,  and the average flight time of the sUAS is increased by 5\%. Since the availability of channels in the environment is limited, a sUAS will not be launched if communication links cannot be established in the flight path. The percentage throughput reduction is less when the traffic becomes lighter.

\begin{figure}[htb] 
  \centering
  \captionsetup{justification=centering,margin=1.2cm}
  \subfloat[$T_{500}$]{\includegraphics[width=1.7in]{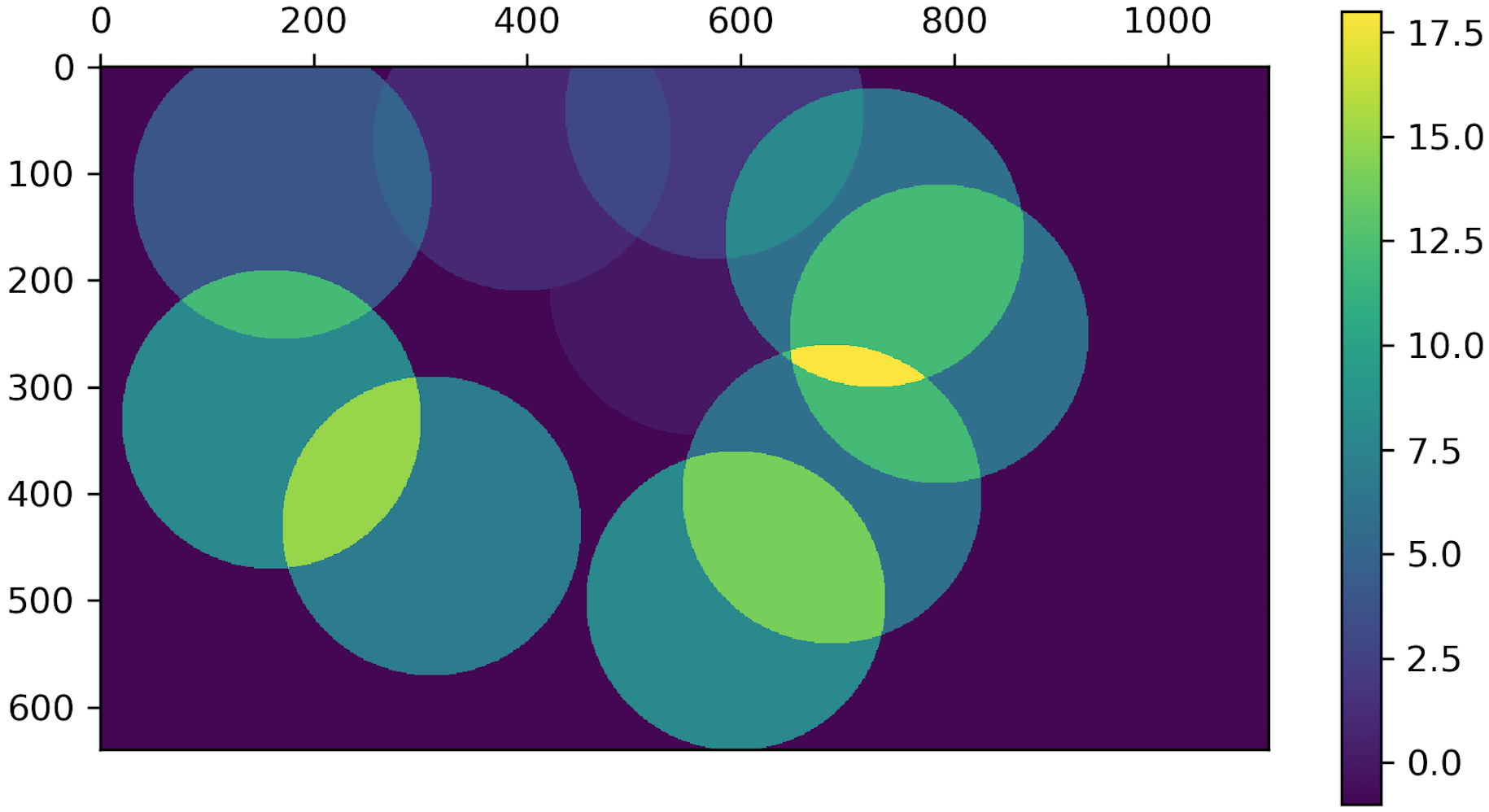}%
  \label{Simulation-T500}} %% label for first subfigure 
  \hfil
  \subfloat[$T_{5000}$]{\includegraphics[width=1.7in]{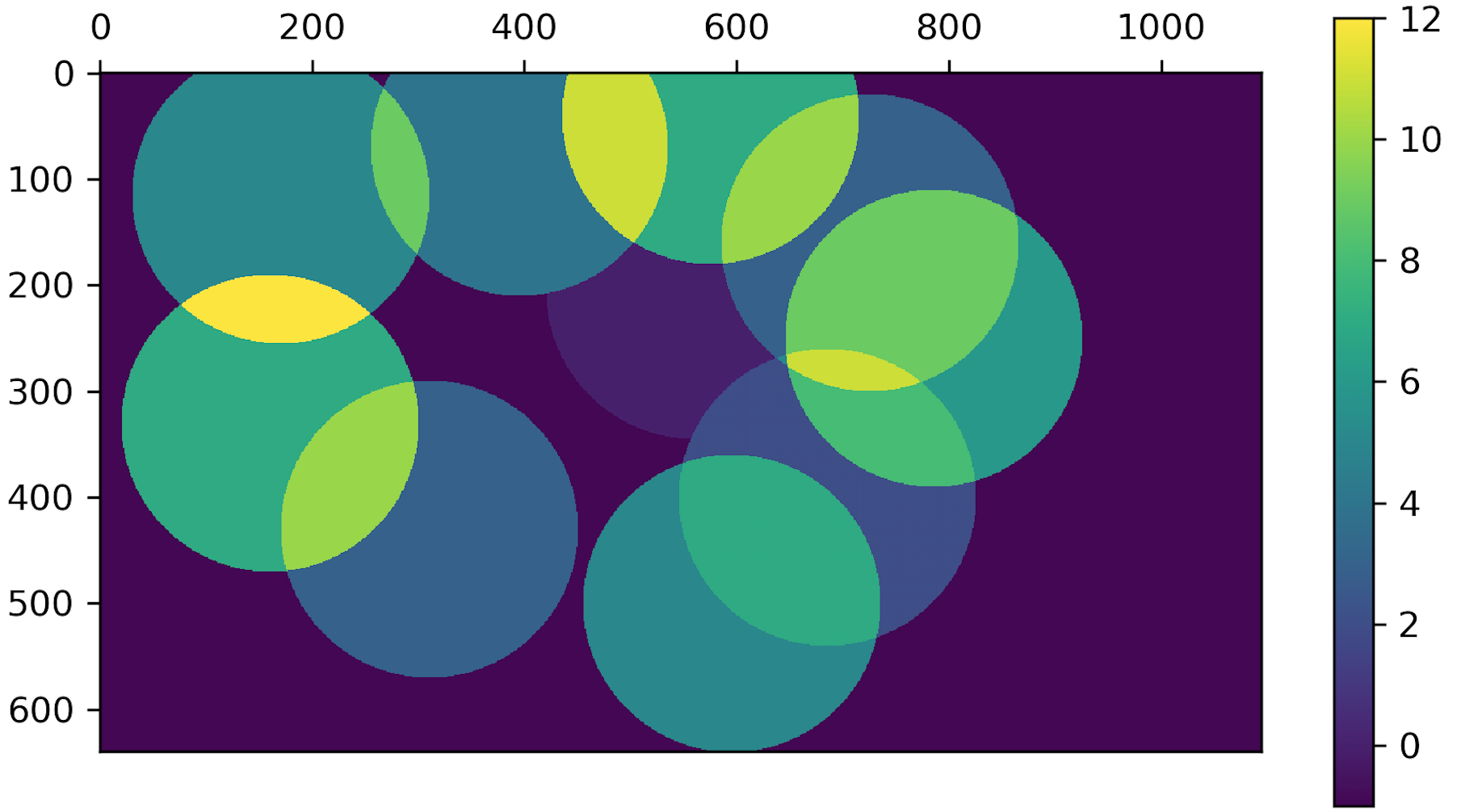}%
  \label{Simulation-T5000}} %% label for first subfigure 
  \hfil
  \subfloat[$T_{10000}$]{\includegraphics[width=1.7in]{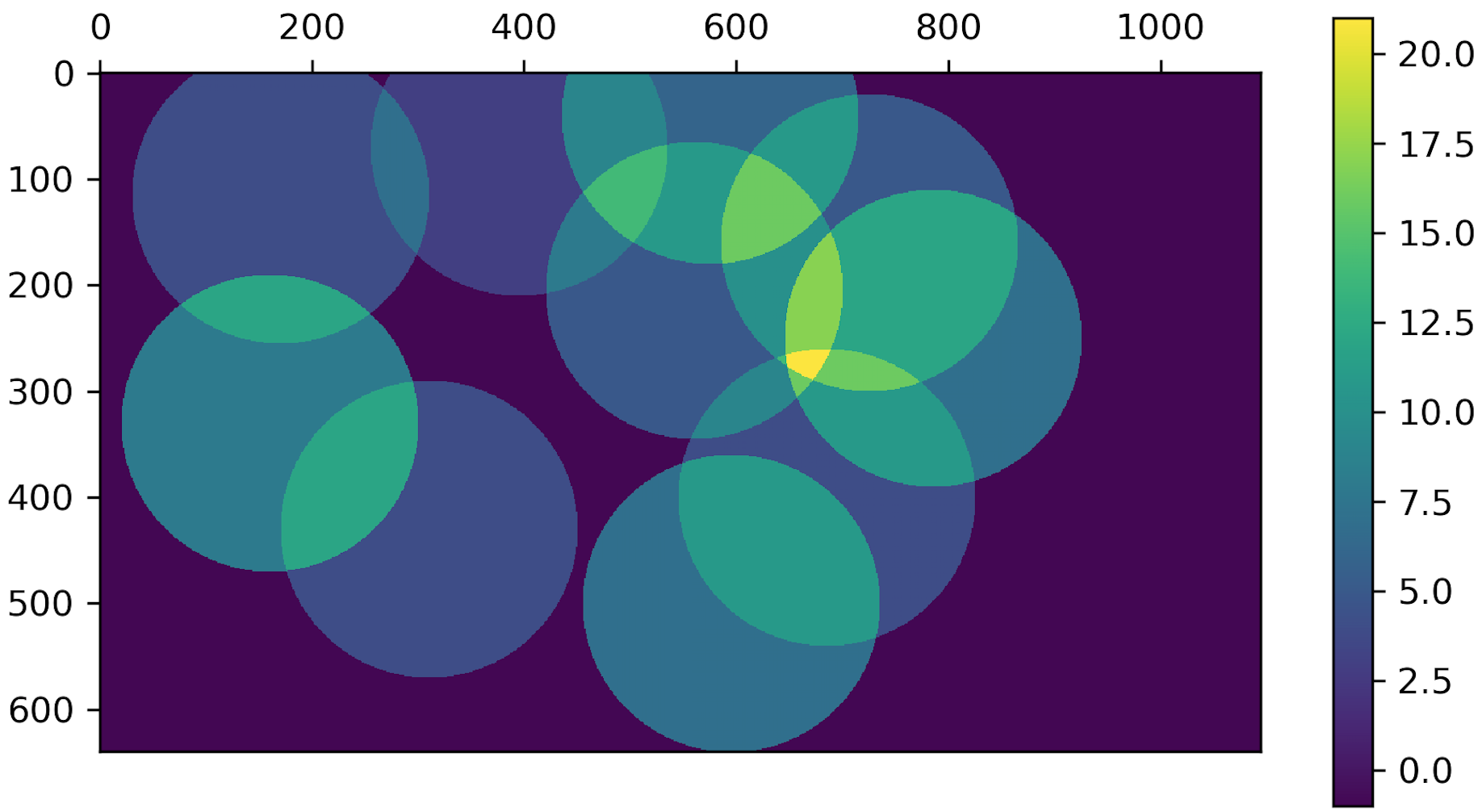}%
  \label{Simulation-T10000}} %% label for first subfigure 
  \hfil
  \subfloat[$T_{19800}$]{\includegraphics[width=1.7in]{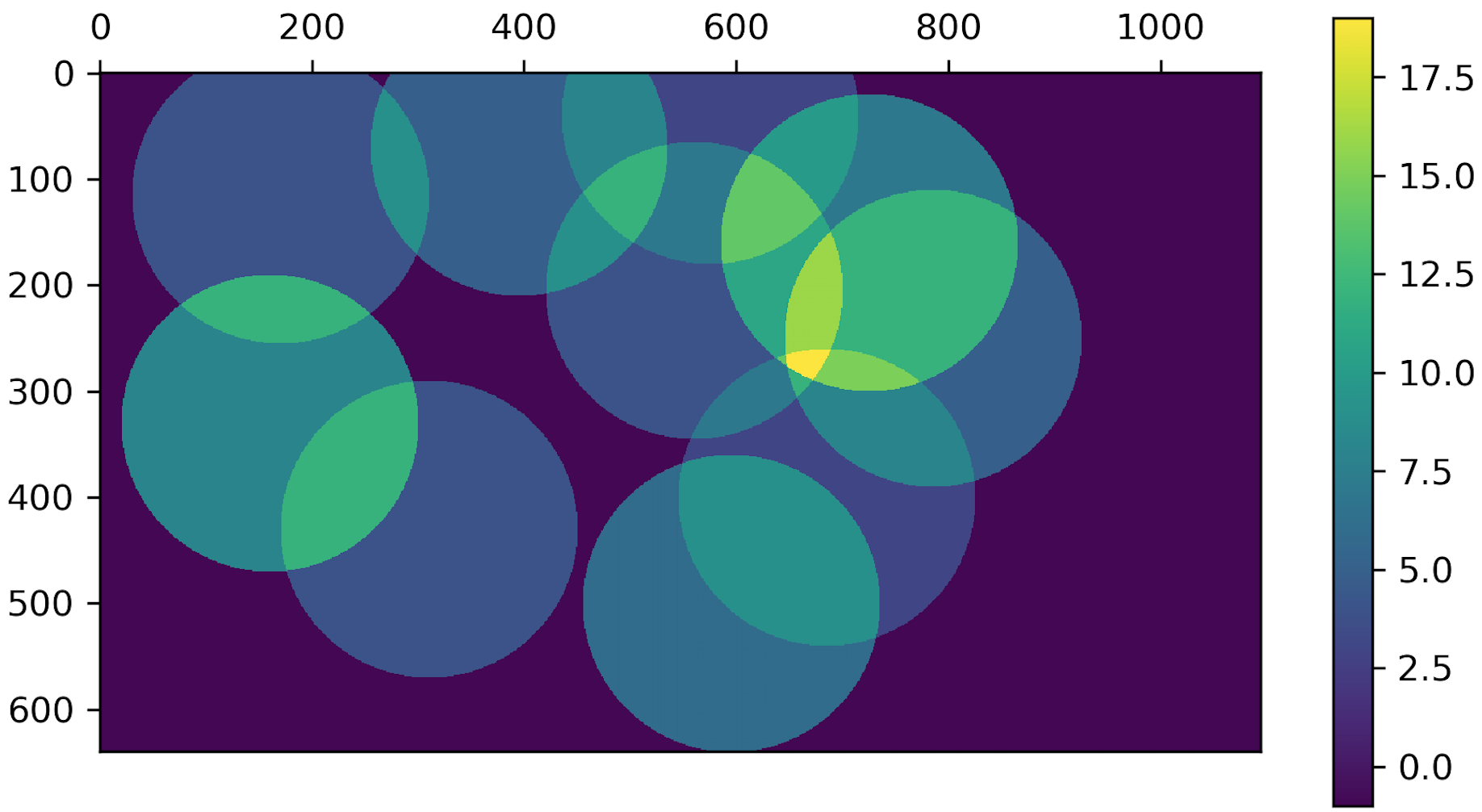}%
  \label{Simulation-T19800}} %% label for first subfigure 
  \caption{Distribution of available cellular channels (Ground Truth)} 
  \label{Connectivity-Simulation} %% label for entire figure 
\vspace{-1em}
\end{figure}

We visualize the resource usage of each base station at some sampled time steps. And the time steps 500, 5000, 10000 and 198000 are chosen. Since the total simulation duration is 20000, the number of airborne UASs at time 5000 and 10000 are representative of the peak value for in-flight UAS in the whole simulation. Fig.~\ref{Connectivity-Simulation} shows the distribution of available communication channels at different time during the simulation. The sample time 500 and 19800 approach to the start time and the end time of the simulation, therefore, the number of in-flight UAS is sparse. The bright yellow represents the available channels are sufficient at that time. However, the deep blue stands for the area where all the communication channels are occupied.

The distribution of available channels in Fig.~\ref{Connectivity-Simulation} are collected from the simulation, hence they represent the “ground truth” information of available communication resources. Based on our observation, the inner belief of the communication resource distribution during the routing stage is very close to the ground truth.  Due to the space limit, we do not plot them here, however, they look just the same as Fig.~\ref{Connectivity-Simulation}. The similarity is expected because the connections between sUAS and the base stations has 90\% of chance to be line-of-sight as mentioned in section~\ref{connectivity}. This means the path-loss is mainly a function of the distance between the sUAS and the base station, and is highly predictable. 

\subsection{Reduction of Control Thrust}
The third experiment compares the control thrust of sUAS trajectories generated using different algorithms. The proactive routing without energy-aware constraint and the reactive “routing” that utilizes artificial potential field \cite{rimon1992exact}  to dynamically avoid potential conflicts are chosen to be the comparison baselines. In proactive routing, we assume that all UASs maintain a consistent speed when they are flying straight. When a sUAS turns, it will initiate a uniform deceleration to slow down and turn 45 degrees, then accelerate uniformly back to its original speed and at the same time complete the turn. 

The Equation\eqref{eqn:acc}  \eqref{eqn:force} are used to estimate the control thrust of an sUAS. Equation\eqref{eqn:acc} calculates the acceleration of the sUAS, $\overrightarrow{v_{t}}$ and $\overrightarrow{v_{t-1}}$ stands for the speed vector at time t and time t-1. Equation\eqref{eqn:force} calculates the force (i.e. control thrust) for the sUAS to follow the trajectory. We assume that the control thrust consists of two parts, the thrust that is needed to maintain a constant speed movement, and the thrust that is needed to accelerate/decelerate. The factor Alpha and Beta are the parameters to scale the force when the UAS fly straight or make turning actions. Large sUAS usually have larger $\beta/\alpha$ ratio.
\begin{equation}
    acc(t) = \overrightarrow {v_{t}} - \overrightarrow {v_{t-1}}
    \label{eqn:acc}
\vspace{-1em}
\end{equation}

\begin{equation}
    force = \alpha \times distance + \beta \times \sum_{0}^{T}|acc(t)|
    \label{eqn:force}
\end{equation}

We vary the value of  and  such that the thrust for the sUAS to make a 90 degree turn with radius 45 meters is similar to the thrust for the sUAS to fly 90, 180 and 270  meters straightly. This corresponding to small, medium and large sUAS respectively. The results from Table \ref{tab3:algorithm_turnpoint} reveal that the proposed algorithm outperforms the two baselines. Especially, the number of turns in our proposed algorithm can be significantly reduced. In order to mitigate the effect of the different flight time, the average energy consumption is normalized by its corresponding flight time. Compared with the no-turn-point-reduction proactive routing method, for the heavy, medium and light traffic, the control trust reductions  are  7.26\%,  7.02\%  and  7.61\%, respectively.  After turn-point reduction, the control trust is similar to the reactive policy for small sUAS and outperforms the reactive policy for large sUAS. We need to point out that sUAS following the reactive deconflict policy does not plan its trajectory with the consideration of communication coverage. Hence it cannot guarantee the safety and connectivity of the sUAS.

\begin{table*}[htbp]
\caption{Control Trust Comparison}
\label{tab3:algorithm_turnpoint}
\centering
\begin{tabular}{|c|*{7}{c|}}
\hline
\multirow{2}{*}{\bf Traffic Type }  &
\multicolumn{1}{c|}{\bf Avg. In-flight UAS } &
\multicolumn{1}{c|}{\bf Avg. Flight Time} &
\multicolumn{1}{c|}{\bf Avg. Turns} &
\multicolumn{3}{c|}{\bf Avg. Energy / Second}\\
& & & &small $\beta/\alpha$ & medium $\beta/\alpha$ &large $\beta/\alpha$  \\
 \hline 
\multicolumn{7}{|c|}{\bf Heavy traffic (generation/10s) }\\
\hline
Proactive w/o. turn point reduction
% & 74.26s & 490.77s &10.57 & 9672.48\\
& 74.26s & 490.77s &10.57 & 19.71 & 21.64 &23.58\\
\hline 
Proactive w. turn point reduction
% & 74.09s  & 487.23s & 1.75 & 8908.99\\
& 74.09s  & 489.62s & 1.82 & 18.21 &18.54 &18.88\\
\hline 
Reactive 
% & 75.75s & 387.64s & 4.01 & 7020.30\\
& 75.75s & 387.64s & 4.01 & 18.11 &18.47 &18.82\\
\hline
\multicolumn{7}{|c|}{\bf Medium Traffic (generation/20s)  }\\
\hline
Proactive w/o. turn point reduction
% & 42.74s & 482.82s &10.12 & 9493.68\\
& 42.74s & 482.82s & 10.12 & 19.66 &21.55 &23.43 \\
\hline 
Proactive w. turn point reduction
% & 43.17s  & 482.83s & 1.68 & 8824.23\\
& 43.17s  & 482.83s & 1.68 & 18.28 &18.58 &18.90\\
\hline 
Reactive 
% & 37.56s & 387.64s & 4.01& 6995.99\\
& 37.56s & 386.54s & 4.01 & 18.05 &18.32 &18.53\\
\hline
\multicolumn{7}{|c|}{\bf Light Traffic (generation/30s)  }\\
\hline
Proactive w/o. turn point reduction
% & 28.04s & 479.35s &10.00 &9421.7\\
& 28.04s & 479.35s &10.00 & 19.66 &21.53 &23.41\\
\hline 
Proactive w. turn point reduction
% & 27.93s  & 476.67s & 1.64 &8710.18 \\
& 27.93s  & 450.11s & 1.65 & 19.35 &19.68 &19.68\\
\hline 
Reactive 
% & 24.23s & 383.39s & 1.60 & 6929.17\\
& 24.23s & 383.39s & 1.60 & 18.07 & 18.16 &18.25\\
\hline
\end{tabular}
\vspace{-1em}
\end{table*}

\subsection{The Resource Usage of the Routing Algorithm}

In the last experiment, we compare the memory usage and computing time of different routing algorithms. Almost all the routing algorithms need to store the environment information of a vast airspace, therefore, there may be a high demand for memory storage during the runtime. We also analyze the time to compute a route for each sUAS. This computation must finish in a very short amount of time so that the launch of the sUAS will not be delayed. Hence, we record the average memory usage and the average routing time during our simulation. The comparison between T-S routing and SRTS routing is given in Table~\ref{tab1:memory}.

\begin{table*}[htbp]
\caption{Routing Algorithm Memory Usage Comparison}
\label{tab1:memory}
\centering
\begin{tabular}{|c|*{7}{c|}}
\hline
\multirow{2}{*}{\bf Traffic Type }  &
\multicolumn{2}{c|}{\bf Heavy Traffic (generation/10s)} &
\multicolumn{2}{c|}{\bf Medium Traffic (generation/20s)} &
\multicolumn{2}{c|}{\bf Light Traffic (generation/30s)} \\
 & 0 No-Fly Zone & 2 No-Fly Zones & 0 No-Fly Zone & 2 No-Fly Zones & 0 No-Fly Zone & 2 No-Fly Zones
\\ \hline 
No Routing & 675MB$\pm$2.2\% & 675MB$\pm$2.2\% & 667MB$\pm$2.1\% & 667MB$\pm$2.1\% & 666MB$\pm$1.5\% & 666MB$\pm$1.5\%\\
\hline 
Baseline T-S Routing & 3513MB$\pm$3.7\% & 3608MB$\pm$3.3\% & 3257MB$\pm$3.1\% & 3315MB$\pm$2.9\% & 2963MB$\pm$3.8\% & 2974MB$\pm$3.5\%\\
\hline 
SRTS Routing & 725MB$\pm$3.4\% & 746MB$\pm$2.3\% & 713MB$\pm$3.8\% & 717MB$\pm$2.7\% & 698MB$\pm$3.2\% & 703MB$\pm$3.0\%\\
\hline
\end{tabular}
\end{table*}

From Table~\ref{tab1:memory} we can see that compared to T-S routing, the SRTS routing method reduces memory usage by more than 70\%. With the help of the instant refreshing mechanism, only the present and future obstacle information will be preserved. And the history obstacle information will been removed automatically. Hence, when increasing from light traffic to medium and heavy traffic, the memory demand of SRTS only increases 2\% and 6\% respectively for scenarios with no-fly zone constraint, and 2\% and 4\% respectively for scenarios without no-fly zone constraint. While the T-S routing’s memory demand increases 11\% and 21\% respectively for scenarios with no-fly zone constraint, and 10\% and 18\% respectively for scenarios without no-fly zone constraint. These number show that SRTS routing is much more scalable than T-S routing in terms of storage complexity. Moreover, the results indicate that including geographical constraints (i.e. no-fly zones) is not a heavy burden for our routing algorithm. Compared with the simulation scenario without the trajectory management, the memory usage of our proposed algorithm increases only about 10.5\%, 7.5\% and 5.5\% in the cases of heavy, medium and light traffics, respectively.

\begin{table*}[htbp]
\caption{Routing Algorithm Running time Comparison}
\label{tab1:speed}
\centering
\begin{tabular}{|c|*{7}{c|}}
\hline
\multirow{2}{*}{\bf Traffic Type }  &
\multicolumn{2}{c|}{\bf Heavy Traffic (generation/10s)} &
\multicolumn{2}{c|}{\bf Medium Traffic (generation/20s)} &
\multicolumn{2}{c|}{\bf Light Traffic (generation/30s)} \\
 & 0 No-Fly Zone & 2 No-Fly Zones & 0 No-Fly Zone & 2 No-Fly Zones & 0 No-Fly Zone & 2 No-Fly Zones
\\ \hline 
Baseline T-S Routing & 2.33ms & 3.20ms & 1.81ms & 2.27ms & 1.49ms & 1.86ms\\
\hline 
SRTS Routing & 0.37ms (-84.12\%) & 0.49ms (-84.69\%) & 0.35ms (-80.66\%) & 0.42ms (-81.50\%) & 0.34ms (-77.18\%) & 0.40ms (-78.49\%)\\
\hline
\end{tabular}
\end{table*}

Moreover, the results from Table~\ref{tab1:speed} show that, by using our SRTS routing algorithm, the UAS route planning time can be significantly reduced. Compared with the original TS routing algorithm, we can achieve 84.69\%, 81.50\% and 78.49\% planning time reduction in the scenarios of high, medium and light traffics, respectively.

\section{Conclusions}\label{sec:conclusion} 

In this paper, we have proposed a new sparse represented temporal spatial routing algorithm for Unmanned Aircraft Systems (UAS) traffic management. The proposed algorithm allows the sUAS to avoid static no-fly areas (i.e. static obstacles) or other in-flight sUAS and areas that have contested communication resources (i.e. dynamic obstacles). The core functionality of the routing algorithm supports the instant refresh of the in-flight environment making it appropriate for highly dynamic air traffic scenarios. In addition, our characterization of the routing time and memory usage demonstrate that our algorithm outperforms a traditional T-S routing algorithm. Finally, the results have shown that the proposed algorithm has the ability to evaluate different sUAS traffic management policies. Moreover, the SRTS routing algorithm can be easily integrated with other simulation tools for further study.

\bibliographystyle{ieeetr}

\bibliography{DASC}

\begin{thebibliography}{10}

\bibitem{zhang2018cellular}
S.~Zhang, Y.~Zeng, and R.~Zhang, ``Cellular-enabled uav communication: A
  connectivity-constrained trajectory optimization perspective,'' {\em IEEE
  Transactions on Communications}, vol.~67, no.~3, pp.~2580--2604, 2018.

\bibitem{azari2019cellular}
M.~M. Azari, F.~Rosas, and S.~Pollin, ``Cellular connectivity for uavs: Network
  modeling, performance analysis and design guidelines,'' {\em IEEE
  Transactions on Wireless Communications}, 2019.

\bibitem{zhao2019simulation}
Z.~Zhao, C.~Luo, Z.~Jin, F.~Basti, M.~C. Gursoy, C.~Caicedo, and Q.~Qiu, ``A
  simulation framework for fast design space exploration of unmanned air system
  traffic management policies,'' {\em arXiv preprint arXiv:1902.04035}, 2019.

\bibitem{puri2005survey}
A.~Puri, ``A survey of unmanned aerial vehicles (uav) for traffic
  surveillance,'' {\em Department of computer science and engineering,
  University of South Florida}, pp.~1--29, 2005.

\bibitem{kopardekar2016unmanned}
P.~Kopardekar, J.~Rios, T.~Prevot, M.~Johnson, J.~Jung, and J.~E. Robinson,
  ``Unmanned aircraft system traffic management (utm) concept of operations,''
  2016.

\bibitem{sastry1995hybrid}
S.~Sastry, G.~Meyer, C.~Tomlin, J.~Lygeros, D.~Godbole, and G.~Pappas, ``Hybrid
  control in air traffic management systems,'' in {\em Proceedings of 1995 34th
  IEEE Conference on Decision and Control}, vol.~2, pp.~1478--1483, IEEE, 1995.

\bibitem{albaker2009survey}
B.~Albaker and N.~Rahim, ``A survey of collision avoidance approaches for
  unmanned aerial vehicles,'' in {\em 2009 International Conference for
  Technical Postgraduates (TECHPOS)}, pp.~1--7, IEEE, 2009.

\bibitem{lavalle1998rapidly}
S.~M. LaValle, ``Rapidly-exploring random trees: A new tool for path
  planning,'' 1998.

\bibitem{bortoff2000path}
S.~A. Bortoff, ``Path planning for uavs,'' in {\em Proceedings of the 2000
  American Control Conference. ACC (IEEE Cat. No. 00CH36334)}, vol.~1,
  pp.~364--368, IEEE, 2000.

\bibitem{tisdale2009autonomous}
J.~Tisdale, Z.~Kim, and J.~K. Hedrick, ``Autonomous uav path planning and
  estimation,'' {\em IEEE Robotics \& Automation Magazine}, vol.~16, no.~2,
  pp.~35--42, 2009.

\bibitem{odelga2016obstacle}
M.~Odelga, P.~Stegagno, and H.~H. B{\"u}lthoff, ``Obstacle detection, tracking
  and avoidance for a teleoperated uav,'' in {\em 2016 IEEE international
  conference on robotics and automation (ICRA)}, pp.~2984--2990, IEEE, 2016.

\bibitem{thanh2018completion}
H.~L. N.~N. Thanh and S.~K. Hong, ``Completion of collision avoidance control
  algorithm for multicopters based on geometrical constraints,'' {\em IEEE
  Access}, vol.~6, pp.~27111--27126, 2018.

\bibitem{wang2015obstacle}
C.~Wang, W.~Liu, and M.~Q.-H. Meng, ``Obstacle avoidance for quadrotor using
  improved method based on optical flow,'' in {\em 2015 IEEE International
  Conference on Information and Automation}, pp.~1674--1679, IEEE, 2015.

\bibitem{li2019autonomous}
Y.~Li, H.~Eslamiat, N.~Wang, Z.~Zhao, A.~K. Sanyal, and Q.~Qiu, ``Autonomous
  waypoints planning and trajectory generation for multi-rotor uavs,'' {\em
  Proceedings of Design Automation for CPS and IoT}, 2019.

\bibitem{eslamiatautonomous}
H.~Eslamiat, Y.~Li, N.~Wang, A.~K. Sanyal, and Q.~Qiu, ``Autonomous waypoint
  planning, optimal trajectory generation and nonlinear tracking control for
  multi-rotor uavs,''

\bibitem{beard2003multiple}
R.~W. Beard and T.~W. McLain, ``Multiple uav cooperative search under collision
  avoidance and limited range communication constraints,'' in {\em 42nd IEEE
  International Conference on Decision and Control (IEEE Cat. No. 03CH37475)},
  vol.~1, pp.~25--30, IEEE, 2003.

\bibitem{song2016rolling}
B.~D. Song, J.~Kim, and J.~R. Morrison, ``Rolling horizon path planning of an
  autonomous system of uavs for persistent cooperative service: Milp
  formulation and efficient heuristics,'' {\em Journal of Intelligent \&
  Robotic Systems}, vol.~84, no.~1-4, pp.~241--258, 2016.

\bibitem{rimon1992exact}
E.~Rimon and D.~E. Koditschek, ``Exact robot navigation using artificial
  potential functions,'' {\em Departmental Papers (ESE)}, p.~323, 1992.

\bibitem{jang2017concepts}
D.-S. Jang, C.~A. Ippolito, S.~Sankararaman, and V.~Stepanyan, ``Concepts of
  airspace structures and system analysis for uas traffic flows for urban
  areas,'' in {\em AIAA Information Systems-AIAA Infotech@ Aerospace}, p.~0449,
  2017.

\bibitem{ferguson2005guide}
D.~Ferguson, M.~Likhachev, and A.~Stentz, ``A guide to heuristic-based path
  planning,'' in {\em Proceedings of the international workshop on planning
  under uncertainty for autonomous systems, international conference on
  automated planning and scheduling (ICAPS)}, pp.~9--18, 2005.

\bibitem{kopardekar2014unmanned}
P.~H. Kopardekar, ``Unmanned aerial system (uas) traffic management (utm):
  Enabling low-altitude airspace and uas operations,'' 2014.

\bibitem{clothier2015structuring}
R.~A. Clothier, B.~P. Williams, and N.~L. Fulton, ``Structuring the safety case
  for unmanned aircraft system operations in non-segregated airspace,'' {\em
  Safety science}, vol.~79, pp.~213--228, 2015.

\bibitem{seet2004star}
B.-C. Seet, G.~Liu, B.-S. Lee, C.-H. Foh, K.-J. Wong, and K.-K. Lee, ``A-star:
  A mobile ad hoc routing strategy for metropolis vehicular communications,''
  in {\em International Conference on Research in Networking}, pp.~989--999,
  Springer, 2004.

\bibitem{nilsson2009quest}
N.~J. Nilsson, {\em The quest for artificial intelligence}.
\newblock Cambridge University Press, 2009.

\end{thebibliography}

\end{document}